\documentclass[twocolumn]{aastex62}

\graphicspath{{./}{figures/}}

\received{\today}
\revised{date}
\accepted{date}
\submitjournal{ApJ}

\shorttitle{Spectral evolution of EX Lup}
\shortauthors{\'Abrah\'am et al.}

\begin{document}

\title{Spectral evolution and radial dust transport in the prototype young eruptive system EX Lup\footnote{Based on observations collected at the European Organisation for Astronomical Research in the Southern Hemisphere under ESO programmes 091.C-0668 and 097.C-0639}}

\author{P. \'Abrah\'am} \affiliation{Konkoly Observatory, Research Centre for Astronomy and Earth Sciences, Hungarian Academy of Sciences, Konkoly-Thege Mikl\'os \'ut 15-17, 1121 Budapest, Hungary} \email{abraham@konkoly.hu}

\author{L. Chen} \affiliation{Konkoly Observatory, Research Centre for Astronomy and Earth Sciences, Hungarian Academy of Sciences, Konkoly-Thege Mikl\'os \'ut 15-17, 1121 Budapest, Hungary}

\author{\'A. K\'osp\'al} \affiliation{Konkoly Observatory, Research Centre for Astronomy and Earth Sciences, Hungarian Academy of Sciences, Konkoly-Thege Mikl\'os \'ut 15-17, 1121 Budapest, Hungary} \affiliation{Max Planck Institute for Astronomy, K\"onigstuhl 17, 69117 Heidelberg, Germany}
  
\author{J. Bouwman} \affiliation{Max Planck Institute for Astronomy, K\"onigstuhl 17, 69117 Heidelberg, Germany}

\author{A. Carmona} \affiliation{Universit\'e de Toulouse, UPS-OMP, IRAP, Toulouse F-31400, France}

\author{M. Haas} 
\affiliation{Astronomisches Institut, Ruhr-Universit\"at Bochum, Universit\"atsstrasse 150, 44801, Bochum, Germany}

\author{A. Sicilia-Aguilar} \affiliation{SUPA, School of Science and  Engineering, University of Dundee, Nethergate, Dundee, DD1 4HN, UK}

\author{C. Sobrino Figaredo} 
\affiliation{Astronomisches Institut, Ruhr-Universit\"at Bochum, Universit\"atsstrasse 150, 44801, Bochum, Germany}

\author{R. van Boekel} \affiliation{Max Planck Institute for Astronomy, K\"onigstuhl 17, 69117 Heidelberg, Germany}

\author{J. Varga} \affiliation{Konkoly Observatory, Research Centre for Astronomy and Earth Sciences, Hungarian Academy of Sciences, Konkoly-Thege Mikl\'os \'ut 15-17, 1121 Budapest, Hungary}
\affiliation{Leiden Observatory, Leiden University, PO Box 9513, NL2300, RA Leiden, The Netherlands}


\begin{abstract}

EX~Lup is the prototype of a class of pre-main sequence eruptive stars defined by their repetitive outbursts lasting several months. In 2008 January-September EX~Lup underwent its historically largest outburst, brightening by about 4 magnitudes in visual light. In previous studies we discovered on-going silicate crystal formation in the inner disk during the outburst, but also noticed that the measured crystallinity fraction started decreasing after the source returned to the quiescent phase. Here we present new observations of the 10~$\mu$m silicate feature, obtained with the MIDI and VISIR instruments at Paranal Observatory. The observations demonstrate that within five years practically all crystalline forsterite disappeared from the surface of the inner disk. We reconstruct this process by presenting a series of parametric axisymmetric radiative transfer models of an expanding dust cloud that transports the crystals from the terrestrial zone to outer disk regions where comets are supposed to form. Possibly the early Sun also experienced similar flare-ups, and the forming planetesimals might have incorporated crystalline silicate material produced by such outbursts. Finally, we discuss how far the location of the dust cloud could be constrained by future JWST observations.  
\end{abstract}



\section{Introduction}
\label{sec:intro}

\begin{figure*}
  \includegraphics[angle=0,width=\textwidth]{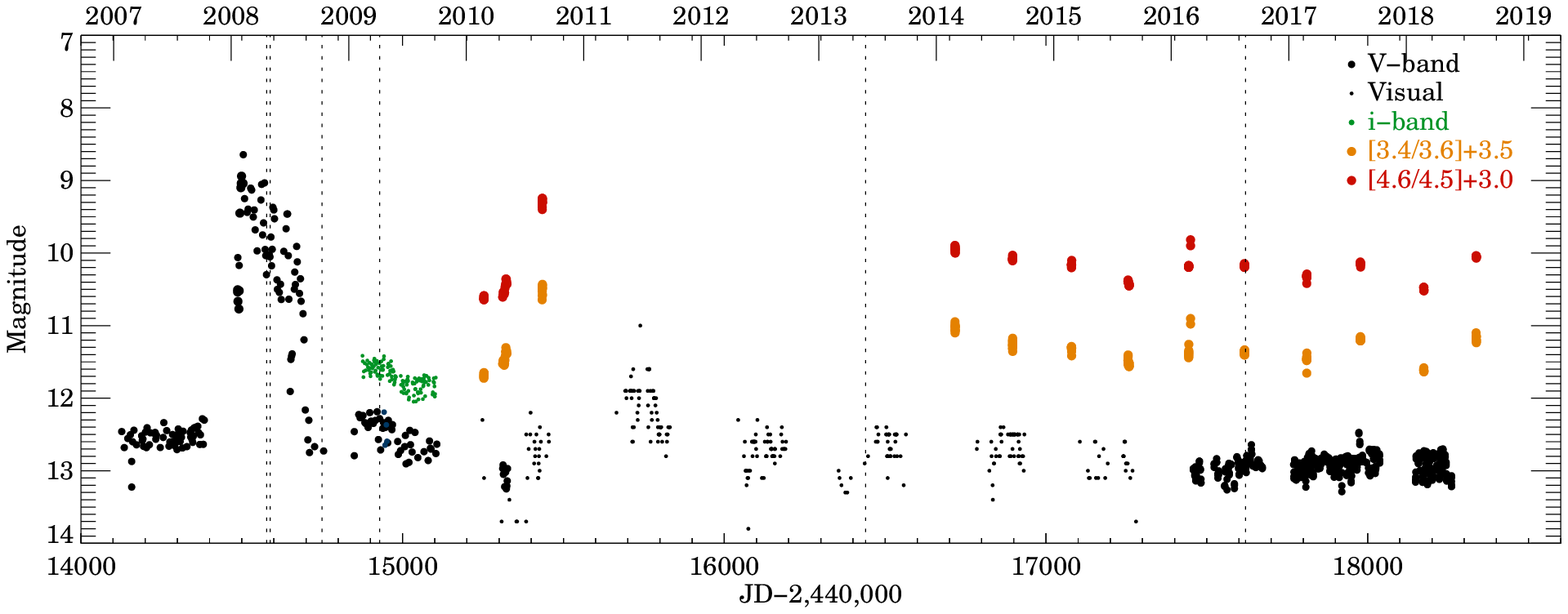}
  \caption{Light curves of EX~Lup. $V$-band observations  before 2010 and visual brightness estimates are from the AAVSO database \href{http://www.aavso.org}{(http://www.aavso.org)}. $V$-band data points from 2016 and later are from the ASAS-SN survey \citep{shappee2014,kochanek2017}. The $i$-band data were observed with the RoBoTT telescope at the Universit\"atssternwarte Bochum (Sect.~\ref{opticalobs}).  The 3.4 and 4.6$\,\mu$m photometry was taken by the WISE satellite \citep{wright2010} and published in the AllWISE Multiepoch Photometry Table and in the NEOWISE-R Single Exposure (L1b) Source Table. Additional 3.6 and 4.5$\,\mu$m data were taken with the Spitzer Space Telescope and were published in \citet{kospal2014}. For better visibility, the infrared light curves are shifted by the values listed in the upper right corner. Vertical dashed lines mark the epochs when mid-infrared spectra were observed (see also Tab.~\ref{tab:log}).}
\label{fig:light}
\end{figure*}

EX~Lup is the prototype of EXors, a class of young pre-main sequence eruptive stars defined by their repetitive outbursts lasting several months \citep{herbig1977,herbig2001,herbig2007}. The flare-ups represent periods of temporarily increased accretion from the circumstellar disk onto the star, possibly in a process similar to the more energetic outbursts of the FU~Orionis-type objects \citep[][]{hk96}. In 2008, EX~Lup underwent its historically largest outburst, brightening by about 4 magnitudes in visual light (Fig.~\ref{fig:light}). Our group obtained a 5 -- 37$\,\mu$m spectrum with the InfraRed Spectrograph (IRS) of the Spitzer Space Telescope on 2008 April 21, shortly after the peak of the outburst. Comparing our spectrum with a similar pre-outburst observation from 2005, we discovered that the originally amorphous silicate grains were transformed to crystalline particles in the inner disk due to the outburst heat \citep{abraham2009}. This was the first direct observation of on-going silicate crystallization in a celestial object. 
The presence of crystalline silicates was confirmed by our VLTI/MIDI interferometric observations obtained in 2008 June and July \citep{juhasz2012}. 

The outburst of EX~Lup ended in 2008 September. In order to document any subsequent changes in the silicate features, we obtained post-outburst Spitzer spectra on 2008 October 10 and 2009 April 6 (Fig.~\ref{fig:spectra}). We expected that the narrow crystalline features in the 10$\,\mu$m emission
(the sharp peak at 10~$\mu$m and the shoulder at 11.3~$\mu$m)
become relatively stronger after the outburst. It is because in the high state both a central crystallized zone, and an outer amorphous region contribute to the mid-infrared silicate emission. In the low luminosity phase, however, the area that is warm enough to emit at 10$\,\mu$m shrinks, and may partly or completely overlap with the zone where surface crystallization had occurred during the outburst \citep[$r<0.7\,$au][]{abraham2009}. But our observations clearly showed that the degree of crystallinity in the 10$\,\mu$m feature {\it decreased} after the outburst, while strong peaks related to cold forsterite grains developed in the 30$\,\mu$m range \citep{juhasz2012}. 

In order to disentangle the effects potentially responsible for the decay of crystalline peaks around 10$\,\mu$m, \citet{juhasz2012} performed radiative transfer and turbulent mixing calculations. Their results excluded that vertical mixing replaced the freshly formed crystals by amorphous particles from the interior of the disk. As an alternative mechanism, they speculated about  outward transportation of the silicate  crystals by a radial wind driven by the outburst. The appearance of forsterite peaks in the 15 -- 30$\,\mu$m range in the late phase of and after the outburst would then be interpreted as the arrival of crystalline particles at larger distances from the star where they cool down and emit at longer wavelengths, supporting the outward radial transport scenario.

The quick drop of crystallinity after an outburst could have been suspected because of an earlier  observational result, too.
 In 1955--56, EX~Lup had already produced a major eruption, with a peak visual brightness of 8.5 magnitude, very close to the maximum of the outburst in 2008 \citep{herbig1977}. The comparable peak luminosities of the two outbursts imply that similar amounts of crystalline silicates must have formed. However, by 2004--2005 (the dates of two pre-outburst Spitzer spectra) the crystalline spectral features vanished, setting an upper limit of 50 years for their disappearance. Aiming to outline the fate of the crystalline  particles after the 2008 outburst and to understand the physical processes in action, here we present new $N$-band observations of EX~Lup and model the disappearance of crystalline particles by their transportation to more distant disk regions. 


\section{Observations and data reduction}
\label{sec:obs}

\begin{table}
\centering
\caption{Log of mid-infrared spectra of EX~Lup analyzed in this paper. EX~Lup underwent a large outburst in 2008 Jan-Aug, thus the first two spectra represent a pre-outburst state, the ones in 2008 Apr-May were taken close to the peak of the outburst, while the remaining spectra are post-outburst observations.}
\label{tab:log}
{\begin{tabular}{l@{}clc}
\hline\hline
Date & Telescope & Instrument/ & AOR/  \\
 &  & Spectral module & Prop. ID \\
\hline
2004 Aug 30    & Spitzer  & ch0, ch1, ch2, ch3 & 5645056   \\ 
2005 Mar 18    & Spitzer  & ch0, ch1, ch3      & 11570688  \\ 
2008 Apr 21    & Spitzer  & ch0, ch1, ch3      & 27039232  \\ 
2008 May 2     & Spitzer  & ch1, ch3           & 27063808 \\ 
2008 Oct 9     & Spitzer  & ch0, ch2           & 28476672  \\ 
2009 Apr 6     & Spitzer  & ch0, ch2           & 28476416  \\ 
2013 Apr 27/28 & VLTI     & MIDI               & 091.C-0668 \\
2016 Aug 20    & VLT      & VISIR              & 097.C-0639 \\
\hline
\end{tabular}
}
\end{table}

\subsection{Mid-infrared spectroscopy with VLT/VISIR}

We obtained $N$-band spectra of EX~Lup using VISIR \citep{lagage2004} on ESO's Very Large Telescope (097.C-0639, PI: P. \'Abrah\'am). The observations were performed on 2016 August 20 with an exposure time of 2300~s, and a slit size of 1$''\times 31{\farcs}74$, which provided a spectral resolution of $R\sim 350$. We observed HIP~78820 as our standard star, but to correct for airmass differences we also used HD~178345, that was measured during the same night  close to the zenith, from the ESO archive. For the basic data reduction and the extraction of the spectra we run the ESO VISIR spectroscopic pipeline. In order to correct for telluric features, we divided the target spectrum by a spectrum derived via interpolation between the two standard star measurements, one obtained at higher and one at lower airmass than EX~Lup. Flux calibration was carried out by multiplying by the model spectra of the standard stars.

\subsection{Archival VLTI/MIDI observations}
\label{MIDIobs}

EX~Lup was observed in the post-outburst phase with the mid-infrared interferometer MIDI \citep{leinert2003}, as part of the program 091.C-0668 (PI: S. Antoniucci).
In total, 17 interferometric observations were carried out on 2013 May 27/28, using the U2-U3 and U3-U4 baselines, with baseline lengths ranging from 31\,m to 62\,m. 
The MIDI data are publicly available, and have been already published in
an on-line interferometric atlas of MIDI observations of low- and intermediate-mass young stars \citep{varga2018}. The reduced data consist of $7.5-13\ \mu$m total and correlated spectra, spectrally resolved visibilities, and differential phases. The visibilities imply that the object was barely resolved at most baselines. Both the total and the correlated spectra show the $10\ \mu$m silicate emission feature. 

Since all MIDI spectra from 2013 exhibited very similar spectral shapes, we averaged them to increase the signal-to-noise ratio. Special care was needed to average correlated spectra, as data taken at different baselines and position angles are generally not comparable. Before the averaging process, we calculated Gaussian sizes from each measurement at $10.7~\mu$m in the same manner as in \citet{varga2018}. 
By combining the resulting sizes with the baseline position angles we could estimate the disk inclination ($62^{\circ +7^\circ}_{\ -20^\circ}$) and position angle ($81^{\circ +4^\circ}_{\ -39^\circ}$). 
These numbers are roughly consistent with $i$=32--38$^{\circ}$ and $PA$=65--78$^{\circ}$, derived from ALMA maps for the outer disk \citep{hales2018}.
Assuming elliptical symmetry, we determined an effective baseline length for each observation as if we would observe a face-on circular disk.  Since in such a configuration the results do not depend on the position angle, we averaged all correlated spectra within two bins of effective baseline of 21--31\,m and 43--51\,m. These average spectra represent disk regions inside $4.1-2.8$~au and $2.0-1.7$~au radii, respectively \citep[adopting the Gaia DR2 distance of 157~pc,][]{brown2018}.
Additionally, we averaged all total spectra, in order to have a spectrum representative of the whole circumstellar disk. 

\subsection{Archival Spitzer spectroscopy}
\label{spitzerobs}

EX\,Lup was observed with the IRS \cite[][]{Houck2004} 
onboard the Spitzer Space Telescope \cite[][]{Werner2004a}
at six epochs between 2004 and 2009 (Tab.~\ref{tab:log}).
Apart from the fourth epoch on 2008 May 2, the
observations were performed using the short-low module ($5.2-14.5\mu$m) of the low-resolution ($R=60-120$) spectrograph.
For the first and the two last epochs also low-resolution spectra using the long-low-module ($14-35\mu$m), were obtained. 
In addition to the low-resolution observations, for epoch one to four, spectra were obtained using the short-high ($9.9-19.5\mu$m) and 
long-high ($18.7-37.2\mu$m) modules of the high-resolution ($R=600$) spectrograph. All observations, with the exception of epoch four, used a PCRS peak-up,
to accurately center EX~Lup at the correct field of view position of the IRS instrument. Observations from four epochs were already published in \citet{juhasz2012}, but here we reprocessed them with the latest calibration files. 

In the case of low-resolution mode the data reduction process started from the \emph{droopres}
intermediate data product. For the high-resolution data we used the \emph{rsc} data product as a starting point. All data products were
processed through the Spitzer Science Center pipeline version S18.18.0.
For the spectral extraction and flux calibration we used the data reduction packages developed for the c2d and feps legacy programs \citep{Lahuis2003, Bouwman2008, Carpenter2008}. 
The background emission in the low-resolution data has been subtracted using associated pairs of imaged spectra from the two 
nod positions, also eliminating stray light contamination and anomalous dark currents. 
Pixels flagged by the data pipeline as being "bad" were replaced with a value interpolated in the dispersion direction 
from an elongated 8~pixel perimeter surrounding the flagged pixel. The spectra were extracted using 
a 6~pixel and 5~pixel fixed-width aperture in the spatial dimension for the Short Low 
and the Long Low modules, respectively. The low-level fringing at wavelengths $>20\mu$m
was removed using the irsfringe package \citep{Lahuis2003}. The spectra were 
calibrated with a position dependent spectral response function derived from IRS spectra and MARCS stellar 
models for a suite of calibrators provided by the Spitzer Science Center. To remove any 
effect of pointing offsets in the short-low module data, we matched orders based on the point spread function of the IRS 
instrument, correcting for possible flux losses \citep[][]{Swain2008}. 
We estimate the accuracy of our flux estimates to be at the level of 1~\%.

For the high-resolution data we applied an optimal source
profile extraction method which fits an analytical PSF derived from sky-corrected calibrator data
and an extended emission component, derived from the cross-dispersion profiles of the
flat-field images, to the cross-dispersed source profile. It is
not possible to correct  for  the  sky  contribution  in  the  high-resolution spectra,
subtracting  the two nod positions as  with the low-resolution observations, due to the small slit length.
The fourth epoch observations had a dedicated background observation which we used to subtract the background emission.
For the other high-resolution observations, we used the background estimate from the source profile fitting extraction
method to remove the background emission. For correcting “bad” pixels we used the irsclean
package. We further removed low-level ($\sim$1\%) fringing using
the irsfringe package. The flux calibration for the high-resolution spectrograph has been done  
in  a similar way as for the low-resolution observations.  For the  
relative spectral response function, we also used MARCS stellar models and 
calibrator stars provided through the SSC. The spectra of the calibration 
stars were extracted in an identical way to our science observations 
using  both extraction methods. As with the low-resolution observations,  
we also corrected for possible flux losses due to pointing offsets. We estimate the absolute flux 
calibration uncertainty for the high-resolution spectra to be $\sim$3\%, slightly higher than that of the low-resolution 
observations. Given that the flux calibration of the short-low module of the IRS spectrograph is the most precise,
for those epochs where these were available, the flux of the high-resolution data was scaled to the flux of the low resolution spectra. 
All scaling factors are consistent with the expected uncertainties. 

\subsection{Optical photometric monitoring}
\label{opticalobs}

EX Lup has been monitored in 2009 between February 12 and September 28
with the 15 cm RoBoTT telescope  at the Universit\"{a}tssternwarte Bochum
near Cerro Armazones\footnote{http://www.astro.ruhr-uni-bochum.de/astro/oca}. The observations were carried out
in the $i$-band ($\lambda_{eff} = 752\,$nm, zero-magnitude flux $f_0 = 3631.0$ Jy).
The good weather conditions allowed us for almost a daily data sampling,
in total there were 134 observing nights.
The light curve of EX Lup is constructed by using 20 calibration stars 
on the same exposures.
The absolute photometric calibration was achieved by comparing EX Lup
to Landolt standard fields which were observed during the same period
of time with the same telescope. 
More details on the telescope, observations and reduction steps 
can be found in \citet{Haas2012}. 


\section{Results}
\label{sec:res}

\subsection{Brightness evolution of EX~Lup}
\label{lightcurve}
In order to check whether the weakening of the crystalline silicate features could be related to changes in the irradiation of the disk, i.e., in the luminosity of the central star, we constructed optical and mid-infrared light curves of EX~Lup between 2007 and 2018 (Fig.~\ref{fig:light}). The data imply that no other eruption of amplitude similar to the one in 2008 has occurred in this period. Localized brightness fluctuations, however appeared several times. 

A remarkable brightness variability pattern can be seen in early 2009, a few months after the end of the large outburst, in both the $V$ and $i$ bands. It began with a deep dimming in late 2008, followed by a rapid brightening of $\sim$0.5~mag in the $V$ band, and then by a gradual fading until 2009 autumn. The brightness increase corresponds in time to a moderate rise of the accretion rate (a factor of few to 10 with respect to the minimum quiescence value) detected spectroscopically by \citet{siciliaaguilar15}. Due to the lack of multiwavelength measurements, it is unclear whether a drop in extinction along the line-of-sight had also contributed to this brightness increase. The gradual fading during 2009 is documented in two optical bands (Fig.~\ref{fig:light}), and the relationship between these two light curves could be fitted with a first order polynomial as ${\Delta}i\sim(0.67\pm$0.06$)\times$${\Delta}V$. The slope term agrees, within the formal uncertainties, with the variability amplitude ratio predicted by the interstellar extinction law for these two wavelengths \citep[0.65,][]{cardelli1989}. Thus the available observations cannot exclude that both accretion and extinction processes have been involved in the variability processes in 2009, but the data do not allow to determine their respective contributions. Variable extinction could have been caused by some rearrangement of the inner disk structure as a consequence of the outburst. 

There are indications for a  brightening also in early 2011, but it is less documented, having only a few infrared data points on the rising part and only visual estimates close to the peak. Simultaneous spectroscopy also reveals a slight increase in the accretion rate during this date \citep{siciliaaguilar15}, which nevertheless stays more than one order of magnitude below the 2008 outburst accretion rate value. Similar short-term variations in the accretion rate of EX~Lup are found throughout its observed history \citep[e.g.][]{lehmann95}.
Apart from these two events, EX~Lup exhibited no obvious brightness variations in the last decade. 


\subsection{Mid-infrared spectroscopy}
\label{sec:MIR}

\begin{figure*}
  \includegraphics[angle=0,height=0.7\textheight]{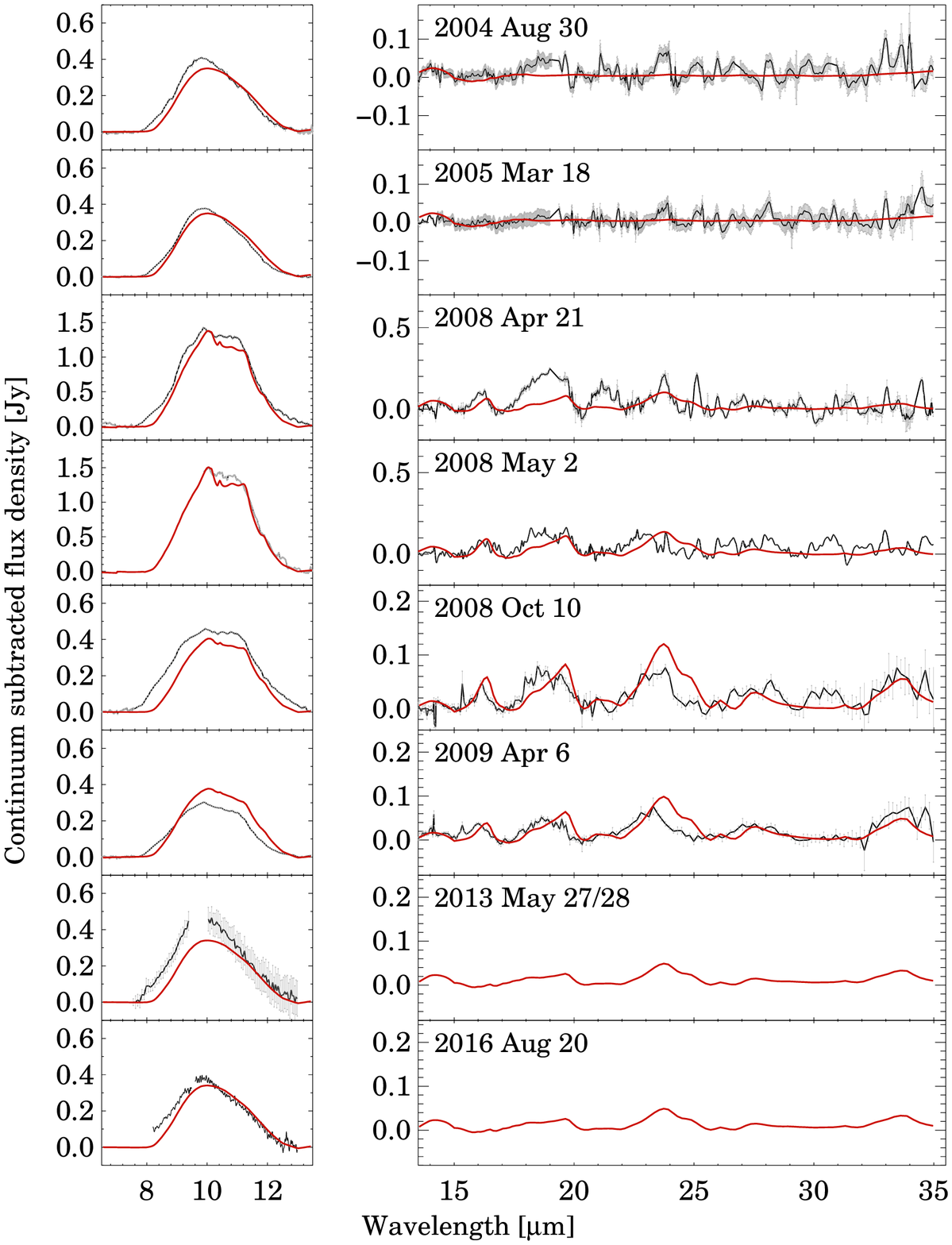}
  \includegraphics[angle=0,height=0.7\textheight]{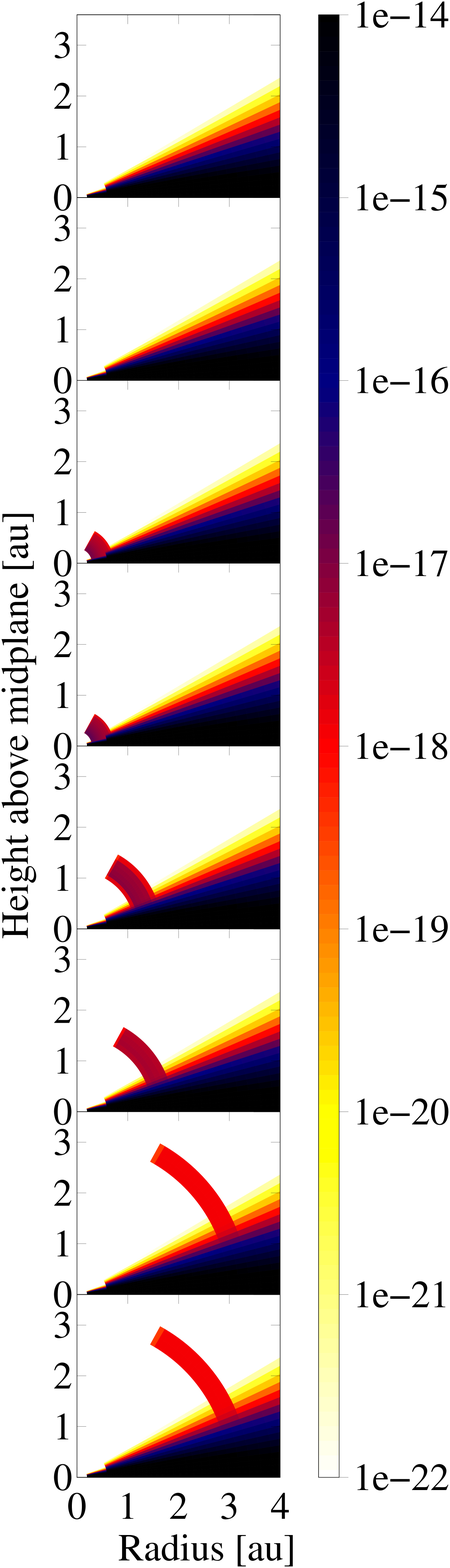}
  \caption{Left and center: continuum subtracted spectra of EX~Lup at different epochs. Black curves are observations, red curves show our models.
  Right: density distributions (unit: $\mathrm{g/cm}^{3}$) of the EX~Lup system in the modeled epochs.
  }
\label{fig:spectra}
\end{figure*}

The list of mid-infrared spectroscopic observations available for EX~Lup is presented in Tab.~\ref{tab:log} (we did not include the VLTI/MIDI observations from 2008 June and July, since the present study focuses on the post-outburst phase and the outburst period is well represented by the higher signal-to-noise Spitzer spectra). Our new VISIR spectrum from 2016, together with the average total spectrum of the MIDI observations from 2013, are plotted in Fig.~\ref{fig:spectra}. For comparison, we also overplotted all Spitzer spectra. Following \citet{juhasz2012}, we subtracted a spline fit continuum from each Spitzer spectrum. The VISIR and MIDI observations included only a short continuum on both sides of the 10~$\mu$m silicate feature, thus we subtracted a simple linear trend. 

In Fig.~\ref{fig:spectra} both the VISIR and MIDI spectra exhibit very similar spectral shapes. The peaks and shoulders that were prominent in the Spitzer spectra due to the presence of crystalline silicates are not visible in these two recent ground-based spectra. Their spectral profiles are reminiscent of the pre-outburst Spitzer observations in 2004 and 2005. The spectrum from 2005 March 18 was modeled by \citet{sipos2009}, who concluded that the emitting dust consisted of small amorphous silicates. Thus, here we may conclude that by 2013, and also later in 2016, any signature of the crystalline forsterite grains formed in 2008 had disappeared from the 10~$\mu$m feature of EX~Lup.

Due to the different baseline combinations, the MIDI interferometric observations offer a possibility to compare the silicate spectra in three different radial zones in the post-outburst period. We found that all the total, and the two correlated spectra (representative of disk regions at $r<2$ au and $r<4$ au, respectively, Sect.~\ref{MIDIobs}),  exhibit $10~\mu$m silicate features characteristic of pristine, amorphous interstellar grains. While in 2008 the MIDI correlated spectra still outlined crystalline spectral features \citep{juhasz2012}, forsterite grains disappeared from the whole inner disk by 2013. 
 

\section{Discussion}
\label{sec:dis}

Analysing the long wavelength parts of the Spitzer spectra of EX~Lup, \citet{juhasz2012} reported the appearance of crystalline silicate features at 23, 28, and 33 $\mu$m after the end of the outburst (see the  2008 October 10 and 2009 April 6 panels in Fig.~\ref{fig:spectra}). This was a strong argument against the post-outburst destruction of the crystals, either by amorphization due to high energy radiation or by being accreted into the star. The result was also not compatible with the vertical mixing scenario, because if the fresh crystals were rapidly mixed down below the disk surface then the appearance of any long wavelength forsterite peak would also not be expected.
Thus, \citet{juhasz2012} proposed a qualitative scenario in which the crystallized grains are transported outward, and after cooling down they emit radiation primarily at wavelengths longward of 10~$\mu$m. In the following we will adopt and perform a quantitative study of this hypothesis, by fitting the crystalline features, and determining the mass and actual radial distance of the crystals for the different epochs. 

\subsection{Modeling}
\label{sec:model}

We performed a quasi-static radiative transfer modeling of EX~Lup in its pre-outburst, outburst, and post-outburst states, using the RADMC3D code \citep{radmc3d}. For the pre-outburst (quiescent) phase, we used a disk model based on \citet{sipos2009}, with only minor modifications due to the updated knowledge of disk inclination from ALMA \citep{hales2018}. Following \citet{sipos2009}, we introduced a curved inner rim, approximated by a vertical step in the radial density distribution. More details about the disk model are presented in Appendix~A and in Tab.~\ref{tab:radmc}. For the post-outburst phase, we supplemented the disk with a tenuous axisymmetric cloud above the disk surface. The disk component was fixed to the best-fit quiescent model (Fig.~\ref{fig:spectra}).




Checking the light curves of Fig.~\ref{fig:light}, it is a reasonable assumption that there were no strong flare-ups since 2008 that could have produced new crystals (the size of the region heated above 1000~K has always been smaller than the inner radius of the disk).  To comply with the information that crystallization was an episodic process in a limited period,
we adopted an expanding hollow  geometry for the crystal-rich component, with the following assumptions. We postulate that all crystal grains were created during the outburst, within the radial range between 0.3 au and 0.7 au \citep{abraham2009}. This cloud of crystals started expanding during the outburst, pushed away by a stellar or disk wind (for a discussion on the possible origin of this wind, see Sect.~\ref{sec:trans}) and we assume that the crystalline dust cloud continues expanding after the end of the outburst. We prescribe the cloud structure to have a constant width of $r_\mathrm{out}-r_\mathrm{in}=0.4$~au, where $r_\mathrm{in}$ is increasing with time. We assume that the mass and dust composition of the expanding cloud is constant during the post-outburst phase. The best fitting value of the cloud mass will be determined by our modeling. Therefore, its optical depth, and in turn the fraction of stellar light that the cloud can absorb and re-emit in the infrared, is decreasing approximately as $\propto r_\mathrm{in}^{-2}$. While the adopted geometry is an ad-hoc assumption, the cloud is mainly optically thin, thus its integrated emission does not depend on the actual spatial distribution of the dust particles.
At each epoch of observation we will fit the shape of the 10$\mu$m emission feature and deduce the actual radius of the dust cloud, from which we can constrain the expansion velocity between the epochs. While we will derive some kinematical information, our study is not a dynamical modeling, rather a quasi-static sequence of equilibrium models.

Based on these assumptions, first we fitted the cloud parameters to match the Spitzer spectra on 2008 October 10 and 2009 April 6. In the best-fit models, the expanding shell consists of 95\% forsterite and 5\% amorphous carbon.
The optical properties of amorphous carbon were calculated with distributed hollow
sphere model \citep{min2005},
assuming grain size of 0.1$\,\mu$m,
using complex refractive index from \citet{jager1998}.
For the optical properties of forsterite we used the results from a laboratory test
\citep{koike2003},
following \citet{juhasz2012} and \citet{abraham2009}.
The crystalline mass in the cloud is $1.9\times10^{23}$\,g, equivalent to $3.2\times10^{-5}\,M_\Earth$.
The obtained radius $r_\mathrm{in}$ is $1.2$~au for 2008 October 10, and $1.5$~au for 2009 April 6.

In order to reproduce the observations in 2013 and 2016, we also run models with increasingly larger $r_\mathrm{in}$. We found that it is about $r_\mathrm{in}\geq 3$~au where the crystalline features in the 10~$\mu$m peak fade below the detection limit due to the low temperature of the grains. We also fitted the Spitzer observations during the outburst (2008 April 21 and 2008 May 2), although with some different assumptions. Since the crystal formation was still ongoing at these two dates, we allowed the total forsterite mass to be a free parameter, but fixed the inner radius of the crystalline cloud to $r_\mathrm{in}=0.3$~au. In addition, during the outburst we included an accretion heating term in our model. The best fitting models are plotted in Fig.~\ref{fig:spectra}, together with the actual density distribution.
In 2013 and 2016 we could not determine the exact location of the dust cloud. Thus, in the figure we plotted models with radii of 3~au for both epochs, although these are lower limits only.

\subsection{Transportation of the crystals}
\label{sec:trans}

\begin{figure}
  \includegraphics[angle=0,width=\columnwidth]{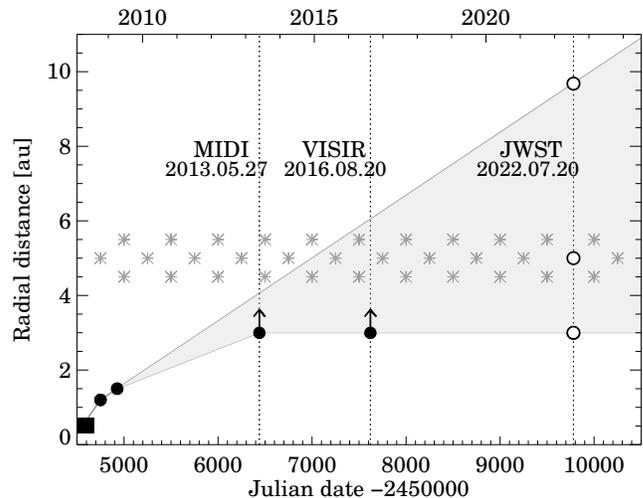}
  \caption{Predicted radial locations of the silicate crystals in the EX~Lup system. 
  Black rectangle marks the time interval of the outburst (x-axis) and the 0.3--0.7~au zone where crystallization  occurred during the outburst (y-axis). 
  Filled black dots are the Spitzer, MIDI and VISIR observations. 
  For the arbitrarily selected epoch of a future JWST observation three possible radii of the expanding dust crystalline cloud are marked. Asterisks denote the approximate location of the snowline.}
\label{fig:trajectory}
\end{figure}

Our modeling suggested that during the $\sim$6 months that separated the two post-outburst Spitzer measurements (2008 October 10 and 2009 April 6), the expanding cloud moved from $r_\mathrm{in}=1.2$\,au to 1.5\,au. This corresponds to an expansion velocity of $\sim$3\,km\,s$^{-1}$. If we assume that after the end of the outburst the crystalline dust cloud was not further accelerated, a higher threshold for the actual location of the cloud is set by a constant velocity expansion of 3\,km\,s$^{-1}$ as marked with the straight line in Fig.~\ref{fig:trajectory}. For a lower threshold, we take into account that after 2013 the cloud must have been at least 3\,au from the star (this result, derived from the spectral fitting, is also supported by our interferometric observations in Sect.~\ref{sec:MIR}). 
The likely position of the inner radius of the expanding forsterite cloud is delimited by the two curves (shadowed area). Note that quadratically adding the Keplerian orbital velocity of the cloud at 1.2 au to the 3\,km\,s$^{-1}$ radial expansion velocity the result is $\approx$17.5\,km\,s$^{-1}$, that is significantly lower than the escape velocity of the system at this radius, $\approx$24.3\,km\,s$^{-1}$. It implies that the expanding cloud will not leave the EX Lup system, but the trajectories of the particles will cross the circumstellar disk at some distances.



The best candidate for driving the motion of the silicate crystals from the inner disk outwards is a wind, either from stellar or disk origin. 
Optical spectroscopic observations of EX Lupi in both quiescence \citep{siciliaaguilar15} and outburst \citep{sicilia12} confirm the presence of
wind features. The wind signatures are ubiquitous during the outburst maximum,
appearing not only in H$\alpha$ and the Balmer lines, but also in the Ca II H and K lines and in the Fe II emission lines. At least two wind components with different densities and temperatures are observed, peaking at $\sim -$200 and $\sim -$100 km/s in outburst \citep{sicilia12}, and their strength decays quickly as the accretion rate decreased. The presence of wind in quiescence is less ubiquitous, and manifests itself mostly in the H$\alpha$ and Ca II H and K lines, which show rapidly variable absorption features at velocities $\sim -$100 to $-$20 km/s that are correlated with the accretion rate, as expected for wind-related features \citep{siciliaaguilar15}. 
The slowest component of the wind is also detected in the Ca II H and K lines. These wind velocities would be nevertheless too high compared to those required for the models, although lacking information on the direction of the wind and considering that EX~Lup is observed at a low inclination angle, the component of
the wind parallel to the disk surface could be significantly smaller than the radial velocity component.

Very low-velocity disk winds have been identified in EX~Lup via  forbidden line emission by \citet{fang18} and
\citet{banzatti19}. These works revealed significant differences in the observed wind velocities in quiescence and in outburst. 
In quiescence, a low velocity component was detected at $-$1.5 km/s \citep[][]{banzatti19}. In outburst, however, velocities were measured at $-$12 km/s \citep[][]{banzatti19} and in the range of $-$15 to $-$18 km/s \citep[][]{fang18}.
Very slow winds and slow outflows have also been detected in FUors \citep{ruiz17}.

The main question is whether these highly variable wind components would be able to induce motion of the silicate grains formed within the disk. While determining the origin and physics of possible wind mechanisms in the EX Lup system is beyond the scope of this paper, there are claims in the literature that winds can have a significant role in radial transportation of grains at distances of 1 au. A possible example 
is the recently proposed scenario for the radial transportation of grains (in particular CAIs) in the early protosolar disk by \citet{liffman16}, which is based on the interaction of the stellar magnetic field with the inner disk rim.

According to Fig.~\ref{fig:trajectory}, by now the crystals are at 3--7 au 
from the central star. From the temperature profile of the EX~Lup disk \citep[][Fig.~4 in Supplementary material]{abraham2009} we can conclude that the water snowline, corresponding to $\sim 160$\,K, is at about 5\,au at the disk surface, and much closer in the midplane. Thus, we suggest that the crystalline particles, produced in the large outburst in 2008, already approached, and in the near future potentially cross the water snowline. If part of them would fall back onto the disk, they could mix with the icy grain population there, and become incorporated into the forming planetesimals, comets. This scenario would help to solve the long-standing mystery of high crystalline fraction observed in pristine comets in the solar system \citep[e.g.][]{hanner94}.

In order to judge the potential impact of an outburst, like that of EX~Lup in 2008, on the mineralogical evolution of the disk, we can estimate the mass of forsterite crystalline particles in our model and compare it with typical comet masses. In our model, the total amount of forsterite in the expanding cloud is $9.5\times 10^{-11}\,M_{\odot}$ or $1.9\times 10^{23}$\,g. As an alternative check of this number, we can compute the expected amount of freshly formed crystals on the disk surface during the outburst.
Assuming that crystallization took place in a region between 0.3 and 0.7 au radii, the surface area of crystal production is $A\approx 1.2$~au$^2$.
We assume that crystallization happened in the hot disk surface where the vertical optical depth $\tau \leq \tau_0 \approx 1$ for the stellar radiation.
Then the vertical column density of this surface layer is $\tau_0/\kappa_{abs}$, where $\kappa_{abs}$ is the absorption coefficient.
Assuming that annealing is efficient in this hot surface and all amorphous silicate grains are turned into crystals, the total crystallized silicate mass in this surface is therefore $M = \tau_0 / \kappa_{abs} \times A \approx \tau_0\times 1.1\times 10^{23}$~g.
This is consistent with the total crystal mass used in our post-outburst models of $\sim 1.9\times 10^{23}$\,g, giving an independent support to the mass estimate in our model, and to the assumption of an optically thin dust cloud.
The crystalline mass of $1.9\times 10^{23}$\,g is equal to $\sim10^4$ times the mass of comet Hale-Bopp, thus the outward transportation of processed crystalline material might contribute significantly to the building blocks of cometary material.

It was a long-standing puzzle that while the extra outburst heat provides optimal conditions for the transformation of pristine amorphous silicate grains into crystalline ones, no sign of crystallization in the outbursts of young eruptive stars was ever seen 
\citep[see e.g. the infrared spectroscopic study of a sample of FUors by][]{quanz2007}. Our infrared spectroscopic observations of EX~Lup's outburst in 2008 revealed crystalline silicates and their formation in a young eruptive star. From that, one would naively expect that a high observed level of crystallinity can be used as a proxy for the erupting nature, and identify dormant eruptive stars in their quiescent phase. However, the results presented in this paper demonstrate that -- at least in an individual case -- crystallinity disappears on a timescale of a few years. Thus, instead of the high level of crystallinity, perhaps it is a rapid temporal variation of the degree of crystallinity that could signal a recent overlooked eruption. Detecting it would require a multiepoch high signal-to-noise ratio spectral monitoring in the mid-infrared of a larger sample of pre-main sequence stars, for which the Mid-Infrared Instrument \citep[MIRI,][]{rieke2015} on-board JWST will be the best instrument in the coming years.

\begin{figure*}
  \includegraphics[angle=0,height=0.3\textheight]{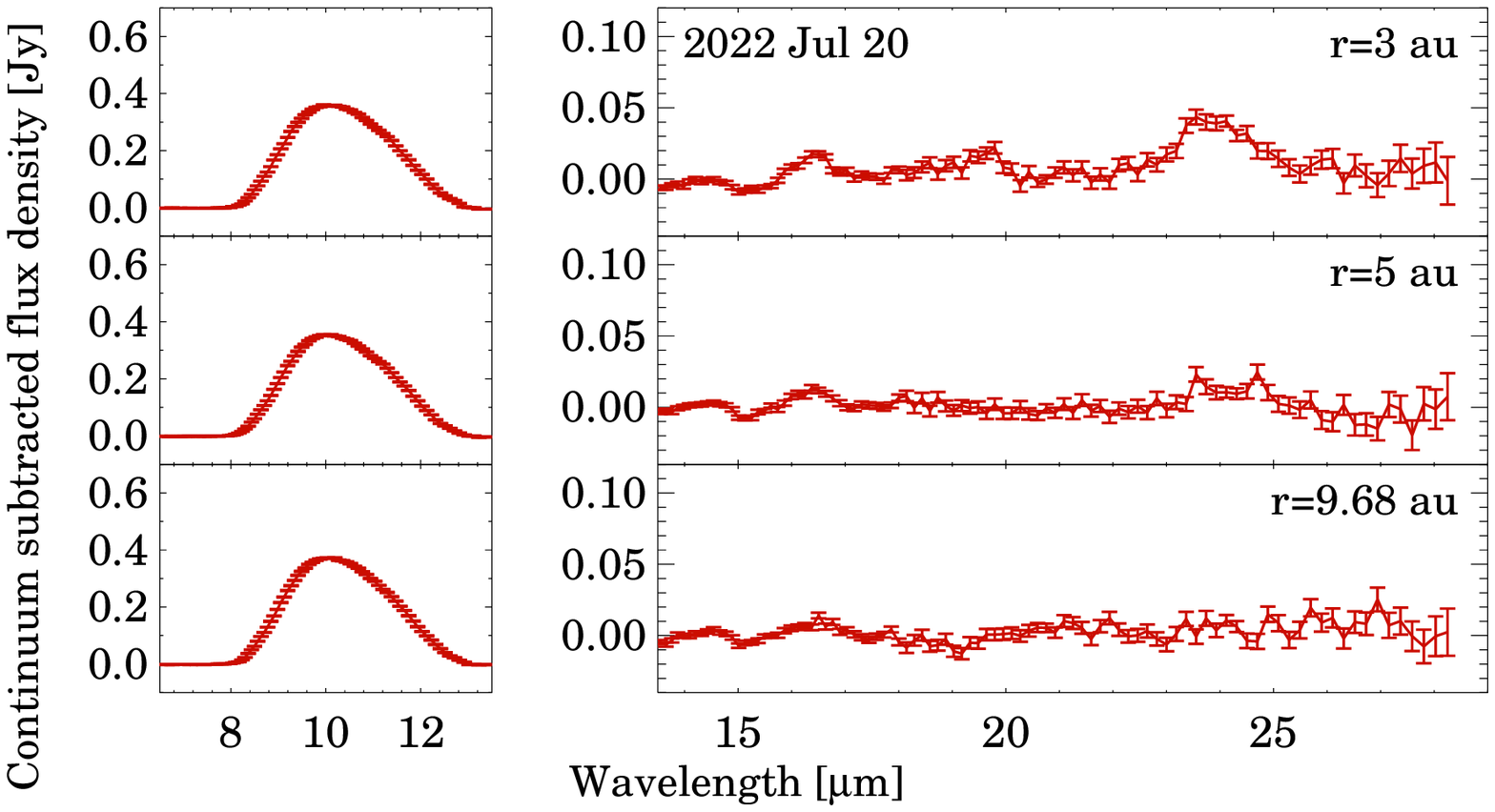} \includegraphics[angle=0,height=0.3\textheight]{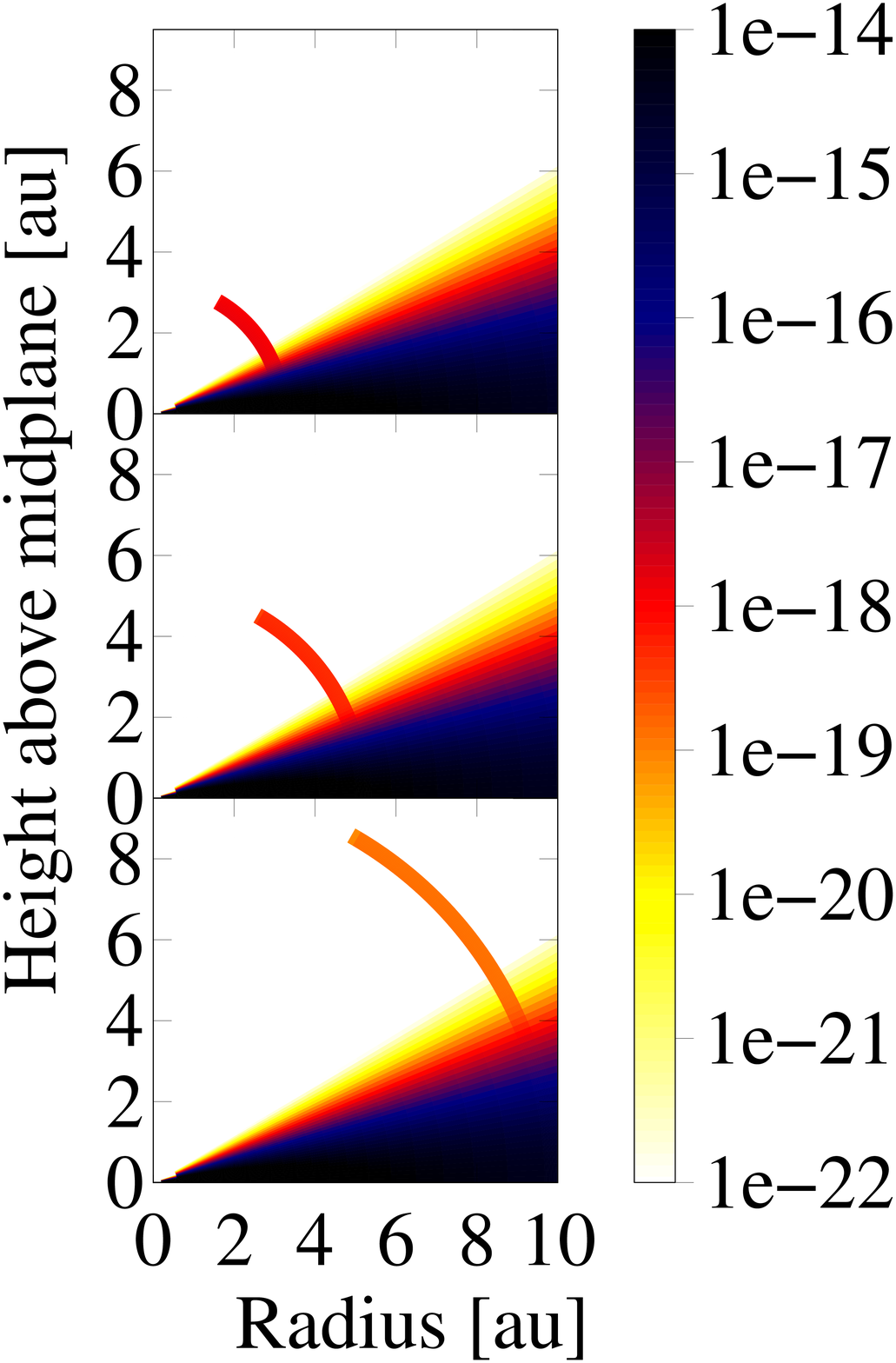}
  \caption{Left and center: continuum subtracted model spectra of EX~Lup at different radii as seen by JWST. 
  Right: density distributions of the EX~Lup system at the modeled radii.
  }
\label{fig:spectra_JWST}
\end{figure*}

\subsection{Detection of the crystalline dust cloud with JWST}

Due to its unprecedented sensitivity, JWST will also be able to observe EX~Lup, and possibly localize the  actual position of the crystalline dust cloud. In Fig.~\ref{fig:trajectory} we marked the epoch of a possible JWST observation set arbitrarily to 2022 Jul 20. The location of the crystalline cloud is unknown at this epoch, thus we calculated the expected spectral shapes for three different radii within the expected range (shaded area in Fig.~\ref{fig:trajectory}). 
The closest possible distance of $r_\mathrm{in}$=3 au is the most favorable case for detection, the intermediate distance at 5 au corresponds to the snowline, and the largest distance at r=9.68 au is consistent with the constant expansion velocity in Fig.~\ref{fig:trajectory}.

We calculated density distributions and integrated spectra for all three radii following the same method as described in Sect.~\ref{sec:model}. Then we used the JWST Exposure Time Calculator V1.3 \citep{pontoppidan2016}, to simulate observations with the MIRI Medium Resolution Spectrograph 
\citep[MRS,][]{wells2015,rieke2015}, assuming an effective integration time of 1000 sec for each of the three dichroic settings, to cover the full wavelength range between 5 and 28$\mu$m. The spectral resolution of the resulting synthetic MIRI spectra was reduced by binning the original full resolution of MRS (R=2000--3000) to be comparable with the the existing EX~Lup observations (R=127). This step increased the S/N of the resulting spectra by a factor of $\sim$20. As a next step, we subtracted a continuum from each of the three spectra in the same way as we did for the Spitzer observations. The resulting spectra and the corresponding density distributions are plotted in Fig.~\ref{fig:spectra_JWST}.

All three model spectra at the different radii exhibit amorphous 10~$\mu$m peaks, which is not surprising as the same was found already at earlier epochs when the dust cloud was still closer to the star (Fig.~\ref{fig:spectra}). However, weak forsterite peaks at longer wavelengths, emitted by cold crystalline grains, may be detectable for JWST/MIRI. From Fig.~\ref{fig:spectra_JWST} we can conclude 
that if the expanding shell is situated at 3~au, then we have a clear detection of the 16, 19 and 24$\mu$m forsterite bands. The intermediate distance at R=5 au seems to be the detection limit for the forsterite bands exhibiting some marginal features, while at 9.68 au none of the bands are visible any more. The results imply that the long-wavelength forsterite features are detectable by JWST up to 3 times larger distances than the radius of the most distant Spitzer observation. If the shell is between 3 and 5 au in 2022, then JWST will determine its actual position, outlining the further path of the expansion motion. A non-detection will probably signal a relatively high expansion velocity and an actual location of $>$5 au. Both results will provide constraints on the effectiveness of the radial mixing in the EX Lup disk.

\section{Summary and Conclusions}
The large outburst of EX~Lup in 2008 crystallized the surface of its inner disk. The strength of the crystalline signatures in the 10~$\mu$m silicate emission peak, however, started weakening soon after the eruption, and our new observations demonstrate that by 2013 no crystalline features were visible in the 10~$\mu$m spectrum. \citet{juhasz2012} claimed that the $>20\mu$m parts of the Spitzer spectra taken in 2008-09 exhibit emission features from cold forsterite grains. The appearance of these long-wavelength indicates that the crystalline particles formed in the outburst were not destroyed by high-energy radiation, not accreted into the star, or not mixed down to the disk midplane, but they could have been transported transported to the outer disk. 

Here we perform a quantitative analysis of this scenario, by fitting at each epoch the spectral shape of the 10~$\mu$m emission feature by a parametric axisymmetric radiative transfer model of an expanding dust cloud of mainly crystalline silicates. Our modeling suggests that in this scenario
the expanding cloud must be at a stellocentric radius larger than 3 au by 2013, roughly corresponding to the snowline radius in the system. A detailed dynamical modeling of the trajectory of the expanding cloud and the paths of the individual crystalline particles is out of the scope of the present paper. Our results, however, suggest that the freshly created crystals could have reached the distance of the water snowline, and if a fraction of them fell back onto the disk they could have been incorporated into forming comets. The crystalline mass of the expanding cloud of 1.9$\times$10$^{23}$\,g ($\sim$10,000 Hale-Bopp-like comets) provides an upper limit for the  amount of crystals transported to the outer disk in a single outburst. 

Our scenario of an expanding cloud could only be validated by new $>20\mu$m observations as the crystals have cooled down by now and radiate at longer wavelengths. We present a simulation of a JWST observation, and demonstrate that with its extreme sensitivity JWST/MIRI will be able to detect the forsterite crystals out to 5 au radius. The successful detection of the expanding motion will support the conclusion that episodic outbursts of young stars may contribute to the build-up of the crystalline dust component ubiquitously seen in comets.



\acknowledgments

This project has received funding from the European Research Council (ERC) under the European Union's Horizon 2020 Research and Innovation programme under grant agreement No. 716155 (SACCRED). This publication makes use of data products from the Near-Earth Object Wide-field Infrared Survey Explorer (NEOWISE), which is a project of the Jet Propulsion Laboratory/California Institute of Technology. AllWISE makes use of data from WISE, which is a joint project of the University of California, Los Angeles, and the Jet Propulsion Laboratory/California Institute of Technology, and NEOWISE, which is a project of the Jet Propulsion Laboratory/California Institute of Technology. WISE and NEOWISE are funded by the National Aeronautics and Space Administration.

\facilities{VLT:Melipal(VISIR), VLTI(MIDI), Spitzer(IRS), OCA:RoBoTT}

\software{ESO VISIR spectroscopic pipeline, Spitzer Science Center pipeline (vS18.18.0), Spitzer irsfringe package 
\citep{Lahuis2003}, RADMC3D \citep{radmc3d}, JWST Exposure Time Calculator V1.3 \citep{pontoppidan2016}}

%






\appendix

\section{pre-outburst disk model}
Our pre-outburst disk model is based on \citet{sipos2009},
with only minor modification due to the updated knowledge of disk inclination \citep{hales2018}.
%
%

For the dust disk, we assumed a power-law radial distribution for its surface density $\Sigma$,
\begin{equation}
\Sigma\left(r\right) = \Sigma_\mathrm{in}\left(\frac{r}{R_\mathrm{in}}\right)^{p}
,~
R_\mathrm{in} < r < R_\mathrm{out}
, \end{equation}
where $\Sigma_\mathrm{in}$ is the surface density at the inner radius $R_\mathrm{in}$, $p$ is the power-law exponent, and $R_\mathrm{out}$ is the outer radius of the dust disk. We assumed the vertical dust density distribution to be a Gaussian function, and the resulting density structure of the disk is
\begin{equation}
\rho \left(r,z\right) = \Sigma\left(r\right)\frac{1}{\sqrt{2\pi}H} \exp \left(-\frac{z^2}{2 H^2} \right),
\end{equation}
where $\rho(r,z)$ denotes the dust density as a function of $r$ and the height $z$ above the mid-plane, and $H$ is the scale height. We assumed that the dependence of $H$ on $r$ is also a power-law, except for the region close to the inner disk rim,
\begin{equation}
h\left(r\right) \equiv \frac{H\left(r\right)}{r} =
\left\{
\begin{array}{lr}
h_\mathrm{rim}, & R_\mathrm{in}<r<R_\mathrm{rim} \\
h_\mathrm{out}\left(\frac{r}{R_\mathrm{out}}\right)^{q}, & R_\mathrm{rim}<r<R_\mathrm{out}
\end{array}
\right.
,\end{equation}
where $h\left(r\right)$ is the dimensionless scale height, $h_\mathrm{in}$ is the dimensionless scale height at the inner radius, and $q$ is the power-law index.

\begin{table*}
\caption{Parameters of the best-fit quiescent model.
Those in italics were kept fixed during the modeling.
}
\label{tab:radmc}
\centering
\begin{tabular}{l p{6.0cm} r l}
\toprule \toprule
\multicolumn{4}{l}{\textbf{System Parameters}}\\
& \emph{Distance} ($d$)                               & 157    &pc\\
& \em{Inclination ($i$)}                    &35      &$^{\circ}$\\
&\emph{Visual extinction} ($A_V$)                    &  0      &\\
\multicolumn{4}{l}{\textbf{Stellar Parameters}}\\
& \emph{Temperature} ($T_*$)       &              3800    &K\\
& \emph{Luminosity}  ($L_*$)                           &0.54     & L$_\sun$ \\
\hline
\multicolumn{4}{l}{\textbf{Disk Component}}\\
& \emph{Inner radius} ($R_\mathrm{in}^\mathrm{d}$)    & 0.3      & au\\
& \emph{Outer radius} ($R_\mathrm{out}^\mathrm{d}$)   &150  &au\\
& \emph{Outer scale height} ($h_\mathrm{out}^\mathrm{d}$)               & 0.12     &\\
& \emph{Power-law index for scale height} ($q$)               & 0.09    &\\
& \emph{Power-law index for surface density} ($p$)               & $-$1    &\\
& \emph{Mass} ($M_\mathrm{dust}$)               & $2.5\times10^{-4}$     & M$_\sun$\\
& Inner rim curvature:\\
& \hspace{0.3cm} Scale height of the rim ($h_\mathrm{rim}$) & 0.05 \\
& \hspace{0.3cm} Ending point of the rim ($R_\mathrm{rim}$) & 0.6  & au\\
\multicolumn{4}{l}{\textbf{Dust composition}}\\
&\emph{Diameter} ($a$) & $0.1$ & $\mu$m \\
&Dust species: & &\\
& \hspace{0.3cm}{amorphous carbon} & 5 & $\%$\\
& \hspace{0.3cm}amorphous olivine & 95 & $\%$\\
\hline
\multicolumn{4}{l}{\textbf{Cloud Component}}\\
& Inner radius ($R_\mathrm{in}^\mathrm{c}$)    	& 1.2 (2008-Oct)      & au\\
&      											& 1.5 (2009-Apr)     & au\\

& \emph{Radial width} ($R_\mathrm{out}^\mathrm{c}-R_\mathrm{in}^\mathrm{c}$)   &0.4  &au\\
& \emph{Center height}$^1$ ($h_\mathrm{center}^\mathrm{c}$)               & 0.6     &\\   
& \emph{Vertical size}$^1$  ($\Delta h_\mathrm{cloud}$)               & 1.2     &\\ 
&Mass ($M_\mathrm{cloud}$)               & $9.5\times10^{-11}$      & M$_\sun$\\
\multicolumn{4}{l}{\textbf{Dust composition}}\\
&\emph{Diameter} ($a$) & $0.1$ & $\mu$m \\
&\emph{Dust species:} & &\\
& \hspace{0.3cm}{amorphous carbon} & 5 & $\%$\\
& \hspace{0.3cm}\emph{forsterite} & 95 & $\%$\\

\hline

\end{tabular}

\flushleft
$^1$ We assume the cloud to extend vertically from
$z/r=h_\mathrm{center}^\mathrm{c}-\Delta h_\mathrm{cloud}/2$
to
$z/r=h_\mathrm{center}^\mathrm{c}+\Delta h_\mathrm{cloud}/2$.

\label{tab:para}
\end{table*}

\bibliography{paper}{}



\end{document}